\documentclass[11pt]{article}
\usepackage{blois,epsfig}

\def\be{\begin{equation}}
\def\ee{\end{equation}}
\def\bea{\begin{eqnarray}}
\def\eea{\end{eqnarray}}

\def\GeVcSq {{\rm (GeV/}c)^2}

\begin{document}
\vspace*{4cm}
\title{RESULTS FROM
PP2PP EXPERIMENT AT RHIC}

\author{ A. SANDACZ }

\address{So\l tan Institute for Nuclear Studies\\
PL-00-681 Warsaw, Poland\\
~~~~\\
\rm{(for the PP2PP Collaboration)}}

\maketitle\abstracts{
We report on the first measurement of the single spin analyzing power ($A_N$) at $\sqrt{s}=200$ GeV, obtained by the pp2pp experiment using polarized proton beams at the Relativistic Heavy Ion Collider (RHIC).  Data points were measured in the four momentum transfer $t$ range $0.01 \leq |t| \leq 0.03$ $\GeVcSq$. Our result is about one standard deviation above the calculation, which uses interference between  electromagnetic spin-flip amplitude and hadronic non-flip amplitude, the source of $A_N$. The difference could be explained by an additional contribution of a hadronic spin-flip amplitude to $A_N$.}

\section{Introduction}
The pp2pp experiment~\cite{PP2PPplb04,lynn,physlett,guryn} at RHIC is designed
to systematically study polarized proton-proton ($pp$) elastic scattering
from \mbox{$\sqrt s =$ 60 GeV to $\sqrt s = $ 500 GeV}, covering
the $|t|$-range from the region of Coulomb Nuclear Interference (CNI)
to 1.5~$\GeVcSq$.
Studies of spin dependence of $pp$ scattering at small momentum transfers and at the highest energies presently available at RHIC offer an opportunity to reveal important information on the nature of exchanged mediators of the interaction, the Pomeron and the hypothetical Odderon (see Ref.~\cite{barone,donnachie} and references therein). The theoretical treatment of small-$t$ scattering is still being developed, hence the experimental data are expected to provide significant constraints
for various theoretical approaches and models (see Ref. \cite{buttimore} and
references therein).

In this paper we present the first measurement of the analyzing power
$A_N$ in $pp$ elastic scattering of polarized protons at RHIC at
$\sqrt{s} = 200 \:\rm{GeV}$ and $0.01 \leq |t| \leq 0.03$ $\GeVcSq$. $A_N$
is defined as the left-right cross section asymmetry with respect to the
transversely polarized proton beam. In this range of $t$, $A_N$ originates
mainly from the interference between electromagnetic (Coulomb) spin-flip and
hadronic (nuclear) nonflip amplitudes \cite{buttimore}.
However, it was realized that $A_N$ in the
Coulomb-nuclear interference (CNI) region is a sensitive probe  of the hadronic
spin-flip amplitude \cite{kz}. A possible hadronic single spin-flip
amplitude would alter $A_N$ and its effect would
depend on the ratio of the single
spin-flip amplitude ($\phi _5)$ to nonflip amplitudes ($\phi _1$ and $\phi _3$), Eq.(~\ref{eq:r5}):
\begin{equation}
r_5 = m \phi _5 / (\sqrt{-t}\:{\rm{Im}}(\phi _1 + \phi _3) /  2 ),
\label{eq:r5}
\end{equation}
where $m$ is the nucleon mass (see Ref. \cite{buttimore} for definitions).

Other measurements of $A_N$ performed at small $t$ have been obtained at significantly lower energies,  by at least a factor of 10,  than the present experiment. These measurements include recent high precision results from the RHIC polarimeters obtained
at $\sqrt{s} = 13.7~\rm{GeV}$  for elastic $pp$ \cite{bravar,jetcal} and
$pC$ \cite{bravar,carboncal} scattering, as well as earlier results from BNL AGS for
$pC$ scattering \cite{e950} at $\sqrt{s} = 6.4~\rm{GeV}$ and from FNAL E704
for $pp$ scattering \cite{e704} at $\sqrt{s} = 19.7~\rm{GeV}$.

The combined analysis of the present result with the earlier ones, especially with the very accurate results of Refs \cite{jetcal,carboncal}, will help to disentangle contributions of various exchange mechanisms involved in elastic scattering in the forward region \cite{trueman}. In particular, such analysis will allow us to extract information on the spin dependence of the diffractive mechanism dominating at high energies.

\section{The experiment and the determination of $A_N$}

\begin{figure}[t]
\begin{center}
\mbox{\epsfig{file=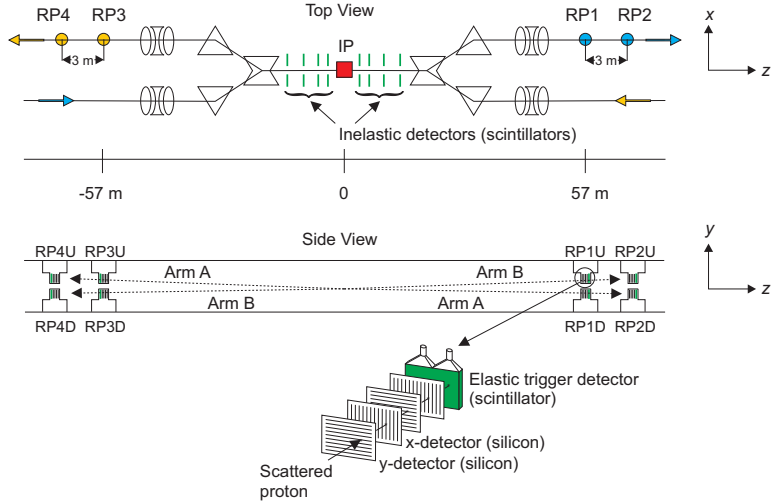,height=3.00in}}
\end{center}
\caption{\footnotesize{Layout of the PP2PP experiment. Note the detector pairs RP1, RP2 and RP3, RP4 lie in different RHIC rings. Scattering is detected in either one of two arms: Arm A is formed from RP3U and RP1D.  Conversely, Arm B is formed from RP3D and RP1U. The coordinate system is also shown.}}
\label{layout}
\end{figure}

The two protons collide at the interaction point (IP), and since the
scattering angles are small, scattered protons stay within the beam pipe of
the accelerator. They follow trajectories determined by the accelerator magnets until they reach the detectors, which measure the $x, y$ coordinates in the plane perpendicular to the beam axis. Those coordinates are measured by Si detectors in the Roman Pots, which are positioned at the location that satisfy so called "parallel to point focusing". More details on the experiment and the technique used can be found in \cite{PP2PPplb04,lynn}.
The layout of the experiment is shown in Fig.~\ref{layout}. The identification of elastic events is based on the collinearity criterion, hence it requires the simultaneous detection of the scattered
protons in the pair of Roman Pot (RP) detectors~\cite{battiston}  on either side of the IP.

To select an elastic event, a match of hit coordinates (x,y) from detectors on the opposite sides of the IP was required to be within $3\sigma$ for x and y-coordinate. The hit coordinates (x,y) of the candidate proton pairs were also required to be in the acceptance area of the detector, determined by the aperture of the focusing quadrupoles located between IP and the RP's. In case that there were more than one match between the hits on opposite  sides of the IP the following algorithm was applied. If there is only one match with number of hits equal to 4, it is considered to be the elastic event. If there is no match with 4 hits or there are more than one such match, the event is rejected.

After selections, the sample of 1.14 million events, for $N^{\uparrow\uparrow}$  and $N^{\downarrow\downarrow}$ bunch combinations, in the $t$-interval $0.010 \le -t \le 0.030$, subdivided into three intervals was used to determine $A_N$.  In each $t$-interval the asymmetry was calculated as a function of azimuthal angle $\phi$ using $5^\circ$-bins. 
The square root formula \cite{ref1} was used for the single spin raw asymmetry
$\varepsilon(\phi)$.
 A cosine fit to the raw asymmetry $\varepsilon(\phi)$ was applied to determine values of $A_N$ (see Ref. \cite{physlett} for more details).

The total systematic error of $A_N$ consists of the scale error of 17.0$\%$ mostly due to the systematic error of the beam polarization measurement, and $8.4\%$ error due to other experimental systematic effects as described in Ref. \cite{physlett}.

\section{ Results and conclusions}
 
\begin{figure}[htp] 
\centering
\begin{tabular}{p{0.45\textwidth}p{0.45\textwidth}}
\epsfig{file=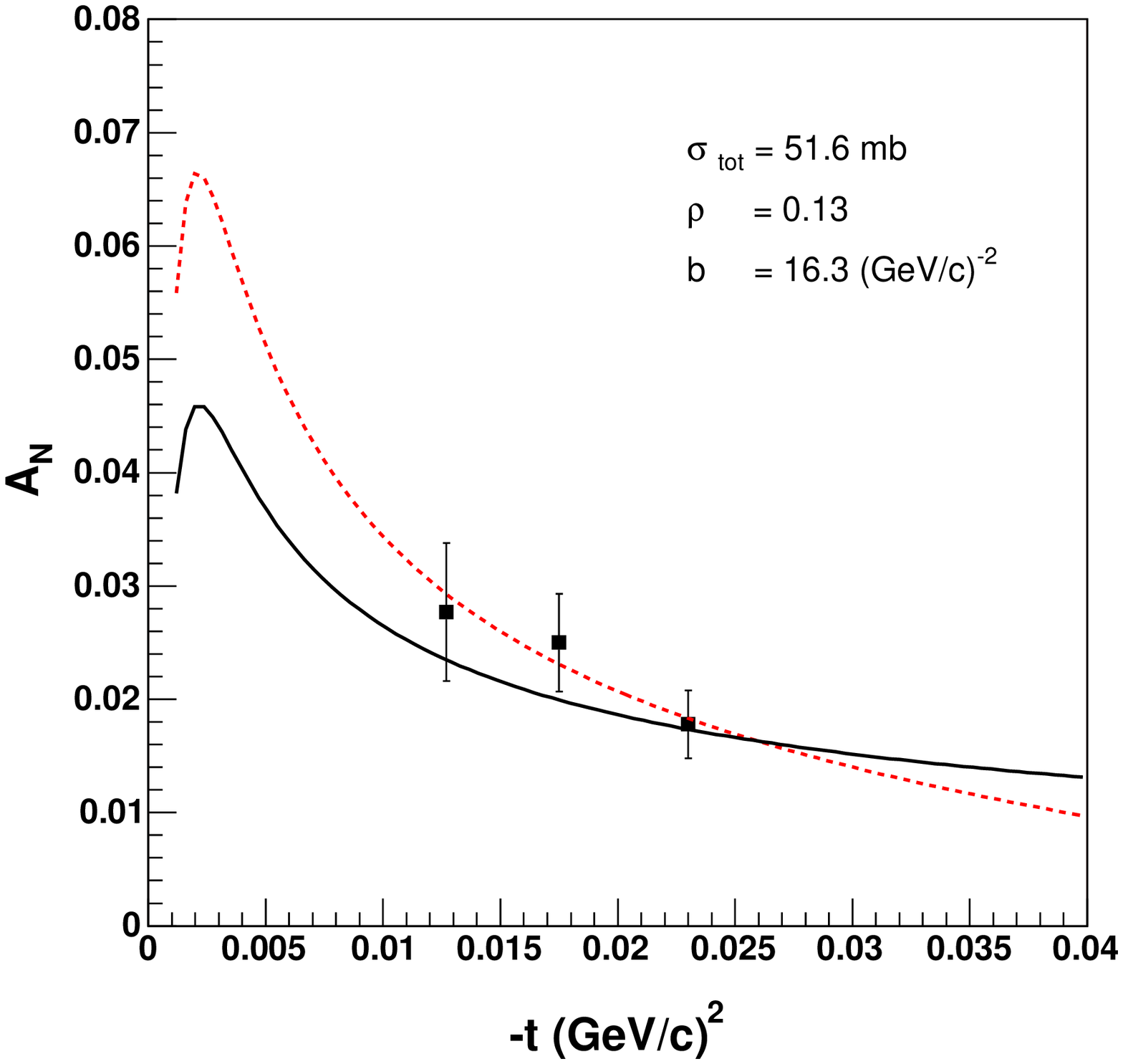,width=0.45\textwidth,clip=} &
\epsfig{file=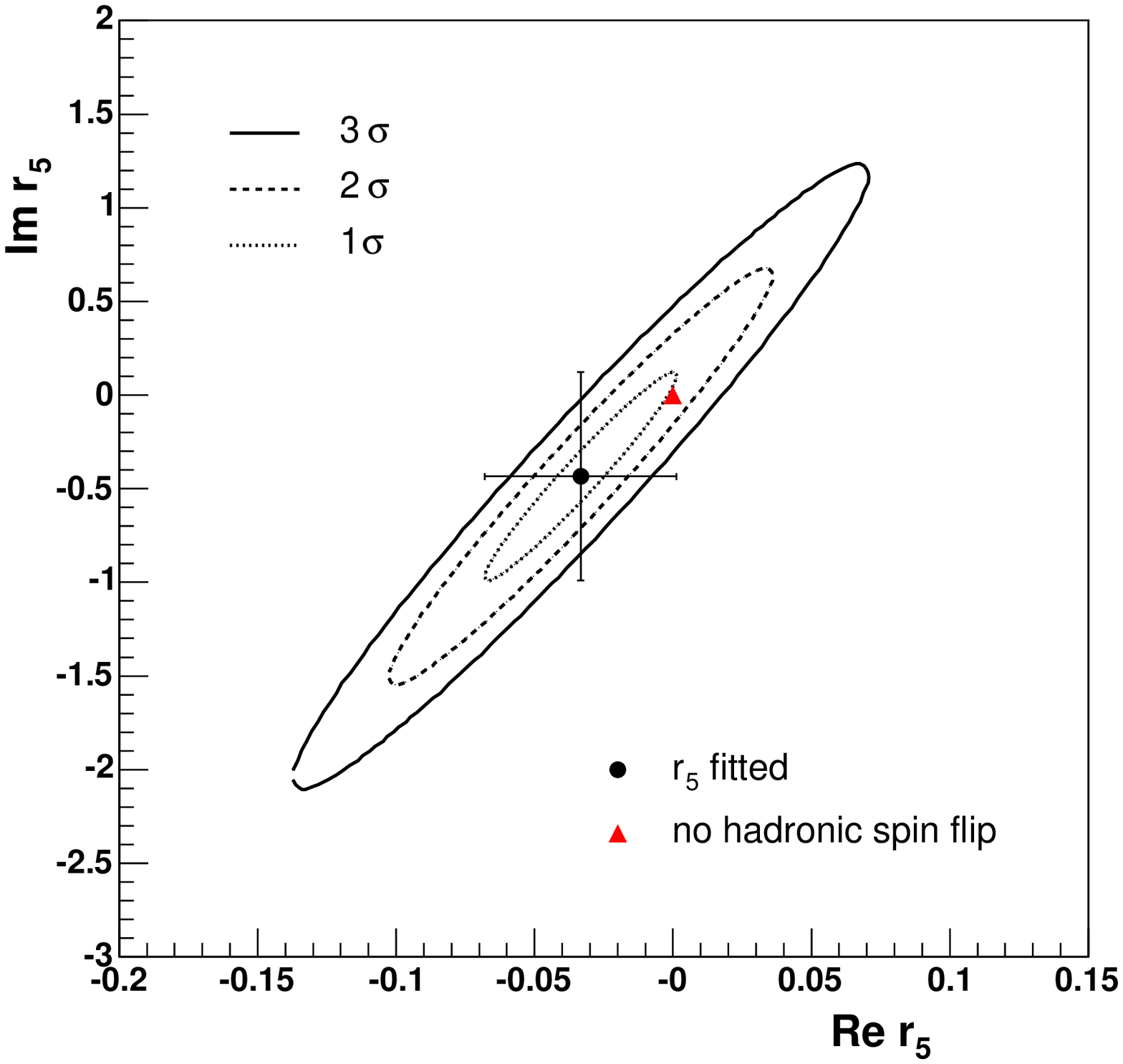,width=0.45\textwidth,clip=}\\
\caption{\footnotesize{The single spin analyzing power $A_N$ for three $t$ intervals. Vertical error bars show statistical errors. The solid curve corresponds to theoretical calculations without hadronic spin-flip and the dashed one represents the $r_5$ fit.\label{fig5}}} &
\caption{\footnotesize{Fitted values of $r_5$ (full circle) with contours corresponding to
the different confidence levels. The point corresponding to no hadronic
spin-flip (triangle) is also shown.}\label{fig6}}
\end{tabular}
\end{figure}

The values of $A_N$ obtained in this experiment and  their statistical errors are shown in  Fig.~\ref{fig5} for the three $t$-intervals.

The curves shown in the figure represent theoretical calculations using the
formula for $A_N$ in the CNI region. The general formula is given by Eq. 28 of
Ref. \cite{buttimore}. With reasonable assumptions that the amplitude $\phi _2$ and the difference
$\phi _1 - \phi _3$ could be neglected at collider energies, the formula
becomes simpler
\begin{equation}
A_N = \frac {\sqrt{-t}}{m} \: \frac {[\kappa (1 - \rho \:\delta) + 2 (\delta \:{\rm{Re}} \:r_5 - {\rm{Im}} \:r_5)] \frac{t_c}{t} - 2 ({\rm{Re}} \:r_5 - \rho \:{\rm{Im}} \:r_5)}{ (\frac{t_c}{t})^2 - 2 (\rho + \delta)\frac{t_c}{t} + (1 + \rho ^2)} .
\label{cnicurve}
\end{equation}
In this formula $t_c = -8 \pi \alpha /\sigma _{tot}$, $\kappa$ is the anomalous
magnetic moment of the proton, $\rho $ is the ratio of the real to imaginary
parts of forward (nonflip) elastic amplitude, and $\delta $ is the relative
phase between the Coulomb and hadronic amplitudes.

The solid curve in Fig.~\ref{fig5} corresponds to the calculation without hadronic spin-flip (${\rm{Re}} \;r_5$ and ${\rm{Im}} \;r_5$ set to 0 in Eq. \ref{cnicurve}). To quantify a possible contribution of the single helicity-flip amplitude $\phi _5$, the formula given by Eq. \ref{cnicurve} was fitted
to the measured $A_N$ values with ${\rm{Re}}\: r_5$ and ${\rm{Im}}\: r_5$ as fit parameters.
The statistical and systematical errors (except the beam polarization error)
of $A_N$ were added in quadrature for the fit.
The results of the fit are following:
${\rm{Re}} \:r_5 = -0.033\pm0.035$ and ${\rm{Im}} \:r_5 = -0.43\pm0.56$. The dashed line
in Fig.~\ref{fig5} respresents the curve resulting from the fit.

The fitted values of ${\rm{Re}} \:r_5$ and
${\rm{Im}} \:r_5$  are shown in Fig.~\ref{fig6} together with contours for
1$\sigma $, 2$\sigma $ and 3$\sigma $ confidence levels. In addition, the point corresponding to no hadronic spin-flip is also shown.
The fitted $r_5$ is compatible, at about one $\sigma $ level, with the hypothesis of no hadronic spin flip. Thus our conclusion is that our results are suggestive of a hadronic spin-flip term, but cannot definitively rule out the hypothesis that only hadronic non spin-flip amplitudes contribute.

\end{document}